\documentclass[showpacs,showkeys,preprint]{revtex4-1}
\usepackage{amsmath}
\usepackage{graphicx}
\usepackage{amsfonts, amssymb}
\usepackage{natbib}

\usepackage{float}
\def\be{\begin{equation}}
\def\ee{\end{equation}}

\def\Hef{\vec{H}_{\mbox{{\tiny eff}}}}

\def\hef{{h}_{\text{\tiny eff}}}
\def\htot{\vec h_{\text{T}}}
\def\hcur{\vec h_{\text{c}}}

\def\grad{\vec \nabla}
\def\kpar{k_u}
\def\kperp{k_d}
\def\eps{\varepsilon}

\DeclareMathOperator{\sech}{sech}
\begin{document}

\title{Domain wall dynamics   for an  in-plane magnetized thin film with large perpendicular hard axis anisotropy  including Dzyaloshinskii-Moriya interaction.}

\author{E.  A. C\'ardenas and  M. C. Depassier}

\affiliation{Instituto de F\'\i sica, Pontificia Universidad Cat\'olica de Chile \\ Casilla 306, Santiago 22, Chile}



\begin{abstract}
We consider a thin ferromagnetic layer to which  an external field  or a current   are applied along  an in plane  easy axis.  The  perpendicular hard axis anisotropy constant is large so that  the out of plane magnetization component is smaller than the in plane components.  A perturbation approach is used  to obtain  the  profile and velocity of the moving domain wall. The dynamics of the in plane components of the magnetization   is governed by a reaction diffusion equation which determines the speed of the profile.  We find a  simple analytic expression for  the out of plane  magnetization  showing  a  symmetric distortion due to the motion  in addition to   the  asymmetric component due to the  Dzyaloshinskii--Moriya interaction.  The results obtained complement previous studies in which either the Dzyalozhinskii vector or the out of plane hard axis anisotropy were assumed small. In the regime studied the Walker breakdown is not observed but the reaction diffusion dynamics predicts a slowing down of the domain wall for sufficiently large magnetic field. The transition point depends on  the applied field, 
saturation magnetization and easy axis anisotropy.

   \end{abstract}


\maketitle

\section{Introduction}

Magnetic domain wall propagation is an active area of research both as an interesting physical phenomenon as well as for its possible applications in logic devices, magnetic memory elements and others \cite{Allwood2005,Stamps2014}.    
 In the micromagnetic approximation    the dynamics of the magnetization  is governed  by the Landau Lifshitz Gilbert (LLG) equation \cite{Landau1935,Gilbert1956}, which cannot be solved except in special cases. The classic Walker solution  takes into account  exchange interaction, interactions modeled as effective anisotropies and studies a domain wall (DW) driven by an external magnetic field. The inclusion of additional physical interactions does not  allow  for simple analytical solutions. Of particular interest and a subject of current research  is the asymmetric Dzyalozhinskii--Moriya interaction (DMI) \cite{Dzyalo1958, Moriya1960}  which leads to new types of  DW inducing a rotation of the  magnetization  and stabilizing chiral  DWs.   Most studies  including DMI address the case of  interfacial DMI in  perpendicularly magnetized thin films  as it  increases significantly  both the DW speed and Walker field \cite{Thiaville2012,Emori2013}. 
 
The role of DMI in in-plane magnetized thin layers has received less attention,  recent  experimental work \cite{Thevenard2017} finds  significant differences with out of plane magnetized  films.  This configuration, including bulk DMI, was studied analytically in \cite{Tretiakov2010, Kravchuk2014, Wang2015, Zhuo2016} using the method of collective coordinates. In  \cite{Tretiakov2010}  as in  \cite{Zhuo2016} the starting point for the application of the collective coordinates method is a profile that neglects the perpendicular hard axis anisotropy which is included as a small perturbation.  A similar approach is taken in \cite{Wang2015} where magnon driven DW motion  is studied.   A different approach is taken in \cite{Kravchuk2014} where the DMI is considered as a  small perturbation .  A fairly complex analytic form for  the perturbation to the static profile due to DMI is found in \cite{Kravchuk2014} and a  linear analysis of this correction to the profile  at the center of the domain wall   provides the  ansatz for the application of the collective coordinate method. The numerical  and analytic results shown in \cite{Tretiakov2010,Kravchuk2014, Wang2015,Zhuo2016} show an asymmetric deformation of the  static DW profile due to DMI. The effect on the speed of different  orientations of the easy axis relative to the Dzyaloshinski vector are studied in  \cite{Wieser2016}.

Here we study a different case.  We are interested in the case where the effective hard axis anisotropy $K_d$ is much larger than the in plane easy axis effective anisotropy $K_u$ and at the same time all the components of the effective field remain of comparable magnitude. These two requirements dictate the scalings needed to perform an asymptotic expansion of the LLG equation using as a small parameter the ratio  of the easy and hard axis anisotropies $K_u/K_d$.   The numerical and analytical work of \cite{Wieser2010,Wang2012,Hu2013,Wang2014} shows that this regime leads to behavior which differs substantially from the the case where  $K_d$ is of the same order as $K_u$.   Materials with  a large disparity of anisotropies were studied  in \cite{Thevenard2012} where  suitably prepared samples of (Ga,Mn)(As,P) with a wide range of effective anisotropies are considered..    In particular,  sample A3  achieves the ratio $K_u/K_d= 0.005$. The speed of propagation of the domain wall in this sample is in qualitative agreement with the numerical results cited above and cannot be explained by the Walker solution. Instability of the Walker solution has been shown analytically in the same scenario \cite{Hu2013}. 
The presence of the small quantity $K_u/K_d$ enables one to  perform an asymptotic expansion of the LLG equation to obtain the leading order dynamical behavior in this regime.  Perturbative approaches to reduce the LLG dynamics to simpler equations with different assumptions have been  employed in \cite{Mikeska1978, GarciaCervera2001, Bazaliy2007, Goussev2016, Lund2016}  among others.  We find that the dynamics of the in-plane components is governed by a reaction diffusion equation and  that the out of plane component  is determined by the in-plane profile and depends on  the applied current and magnetic field in addition to the DMI interaction. The  in- plane components of the magnetization for a tail to tail (TT) configuration are given by the usual profile
\begin{equation}
m_x^{TT} = \tanh \left(\frac{x- v t}{\Delta}\right), \qquad m_y^{TT} = \text{sech}  \left(\frac{x- v t}{\Delta}\right), 
\end{equation}
where the width is given in terms of the exchange constant $C_{\text{ex}}$ and the easy axis anisotropy $K_u$ by   $\Delta = \sqrt{C_{\text{ex}}/(2 K_u)}$.
 
For an external current $u$  and magnetic field $H_a$  applied along the easy $x$ axis, the out of plane magnetization is found to be
\begin{equation} \label{mzi}
m_z^{TT}  =  \frac{M_s}{ 2 |\gamma| \Delta K_d} \text{sech} \left(\frac{x- v t}{\Delta}\right)  \left[   u - v +  \frac{2 D |\gamma| }{M_s}  \tanh \left(\frac{x- v t}{\Delta}\right)  \right],
\end{equation}
where the  speed  $v$  for the TT domain wall in the limit studied is given by
\[
v=  - \mu_0 |\gamma| \frac{\Delta H_a }{ \alpha} + \frac{\beta u}{\alpha}
\]
where  $D$,  $H_a$ and $u$ are  the Dzyaloshinski  constant and the applied magnetic field  and  current respectively.  The quantities $\alpha$, $\beta$  and $\gamma$  are the  Gilbert damping,   the nonadiabatic \cite{Thiaville2005} parameter  and the electron gyromagnetic ratio. An equivalent  expression holds for a head to head (HH) domain wall.  The static profile  is in qualitative agreement with the numerical results reported in \cite{Wang2015} and with the analytic solution for small DMI obtained in \cite{Kravchuk2014}.  The explicit expression for the out of plane magnetization showing the effect of the DMI and applied field and current on the profile for large easy plane anisotropy, Eq.(\ref{mzi}),  has not been reported elsewhere to the best of our knowledge.  The main effect of the motion is to change the symmetry property $m_z(-x) =  - m_z(x)$ of the static solution losing all symmetry as it moves.  The above solution is found  from the leading order expansion of the LLG equation.  

The dynamics in the limit studied is different from that found for the thin film with small perpendicular anisotropy or with small DMI \cite{Tretiakov2010, Kravchuk2014}.
The reaction diffusion equation which governs the in-plane dynamics shows that for sufficiently large external field an initial perturbation will not evolve into this exact analytic solution but it will evolve into a Kolmogorov-Petrovskii-Piscounov  (KPP) \cite{KPP} domain wall moving with slower  speed but qualitatively similar profile. The transition point and this slower speed are given below.  Previous numerical work has shown a  slowdown of the domain wall before encountering the Walker field  \cite{Wieser2010,Wang2012,Wang2014}  for thin films with very large hard axis anisotropy. They attribute the slow down of the domain wall to spin wave emission.  
The analytical results found in this work show similar qualitative behavior  as that found numerically. In the present approach we are able to obtain the envelope of the domain wall, therefore  identification of  the slow down by spin wave emission with a transition from pushed to pulled fronts  is not possible without further work. 

In Section II we state the problem, in Section III we perform an asymptotic analysis and solve the resulting equation and in Section IV we summarize the results.

\section{Statement of the problem}

   We consider a thin narrow film   in the $(x,y)$ plane, with the easy axis $x$ along its length and an effective  hard axis perpendicular to the thin film plane.   A constant external field  and current are applied  along the easy axis $\vec H_a = H_a \hat x$, $\vec u = u \hat x$. \cite{Wang2015}

The material has magnetization $\vec M= M_s \vec m$ where $M_s$ is the saturation magnetization and $\vec m = (m_1,m_2,m_3) $ is the unit vector along the direction of magnetization. 
The dynamic evolution of the magnetization  is governed  by the LLG equation, including the current 
\be\label{LLG}
 \frac{\partial \vec m }{ \partial t} = - \gamma_0  \vec m \times \Hef + \alpha \vec m \times \frac{\partial \vec m }{ \partial t}  - (\vec u \cdot\grad) \vec m + \beta  \vec m \times (\vec u \cdot\grad) \vec m
 \ee
where $\Hef$ is the effective magnetic field,  $\gamma_0=  | \gamma | \mu_0$,   $\gamma$ is the gyromagnetic ratio of the electron,   $\mu_0$ is the magnetic permeability of 
vacuum. The constant  $\alpha>0$ is the dimensionless phenomenological Gilbert damping coefficient. The   vector $u$ is proportional to the current density $j_e$ and has units of velocity. The dimensionless constant $\beta$ is the non adiabaticity parameter.

    The effective magnetic field  is given by  \cite{Wieser2010,Wang2012,Hu2013, Wang2014,Thevenard2012,Kravchuk2014,Wang2015}
\be \label{H1}
\Hef = H_a \hat x + \frac{C_{\rm ex}}{\mu_0 M_s^2}  \frac{\partial^2 \vec M}{\partial x^2} + \frac{2 K_u}{\mu_0 M_s^2} M_1\hat x -  \frac{2 K_d}{\mu_0 M_s^2} M_3 \hat z + \vec H_{\text{DMI}},
\ee
where
$K_u$ the easy axis effective uniaxial anisotropy and $K_d$ is an effective  hard axis anisotropy.  For the effective field due to the exchange interaction we have introduced the constant $C_{ex}$  in terms of which the exchange energy density is written as ${\cal{E}} = \frac{C_{ex}} {2 \mu_0 M_s} |\nabla \vec m|^2$.  This constant is twice the exchange constant $A$ as defined in \cite{HubertSchafer}. 
We have assumed that the demagnetizing field has  a local expression  as an additional  anisotropy in the direction perpendicular to a thin film plane, as demonstrated rigorously in \cite{Gioia1997}.  The combined effect of a local approximation for the demagnetizing field plus crystalline anisotropies and stress induced anisotropies may be represented by effective anisotropies \cite{Thevenard2012}.   We will consider bulk DMI for which the effective field is given by $ \vec H_{\text{DMI}} = - (2 D/ \mu_0 M_s)  \vec{\nabla} \times \vec{m}$.

The effective field (\ref{H1}) has been considered in previous work \cite{Wang2015, Zhuo2016} and has been treated analytically in the case of vanishing or very small $K_d$ or very small $D$. Here we focus on the opposite regime, in which $K_d >>K_u$ and  assume that the effective field due to DMI is comparable to the effective fields due to exchange interaction  and anisotropies.

Our purpose is to obtain a simple analytical description  for the profile of the domain wall  exhibiting explicitly the distortion of the profile due to the Dzyaloshinskii-Moriya interaction and due to the motion of the domain wall. We neglect any possible tilting of the domain wall and assume a  one dimensional model. 
Under such assumption  the magnetization  depends on the easy axis coordinate, $\vec M(x,y,z) = \vec M(x)$ so that the DMI field reduces to 
\be
\vec H_{\text{DMI}} = \frac{2 D}{ \mu_0 M_s}(   \frac{\partial m_3}{\partial x} \hat  y   -  \frac{\partial m_2}{\partial x} \hat  z  ).
\ee

The effect of the current can be expressed as the additional field
\be
H_C = - u  \vec m \times \frac{\partial \vec m}{\partial x} - \beta u \frac{\partial \vec m}{\partial x}.
\ee

Introducing $M_s$ as unit of magnetic field, and introducing the dimensionless space  and time variables   $\xi =  x /L$ with $L= \sqrt{C_{ex}/K_u}$  and $\tau=  t/T$ with 
$T = 1/ (\mu_0 |\gamma | M_s) $  we rewrite equations (\ref{LLG}) and (\ref{H1}) in dimensionless form
\be\label{LLG2}
 \frac{\partial \vec m }{ \partial \tau} = -   \vec m \times \htot + \alpha \vec m  \times \frac{\partial \vec m }{ \partial \tau } 
  \ee
with  $\htot = \hef + \hcur$, 
where 
 \be
\hcur  =  - U  \vec m \times \frac{\partial \vec m}{\partial \xi} - \beta U \frac{\partial \vec m}{\partial \xi},
\ee
and
\be \label{hef}
\hef= h_a \hat x + \frac{1}{2} \kpar  \frac{\partial^2 \vec m}{\partial \xi^2} + \kpar m_1 \hat x - \kperp m_3 \hat z +   d (   m_{3\xi} \hat y -  m_{2\xi} \hat z).
\ee
 Here  $h_a$ is the dimensionless applied field and the dimensionless numbers that have appeared are $\kpar= 2 K_u/(\mu_0 M_s^2)$, $\kperp= 2 K_d/(\mu_0 M_s^2),
 U = u  T/L $ and  $d  = 2 D/(\mu_0 L M_s^2).$
 
 Equation (\ref{LLG2})  together with the expression for the  total field $\htot$ define the problem under study.

\section{Asymptotic development }


We are interested in the regime where $K_u \ll K_d$ so that we expect that the out of plane magnetization $m_3$ will be smaller than the in-plane components, that is,  $m_3 \ll m_1,m_2$. Far from the domain wall we know that the magnetization along the easy axis, $m_1 \rightarrow \pm 1$.  Furthermore,  we wish to consider the situation in which all the components of the effective field are of the same order. In dimensionless variables this implies  $\kpar \ll \kperp$, 
$ \kpar m_1 \sim \kperp m_3$ with $m_3\ll m_1$.   Denoting by $\epsilon$ the size of the ratio $\kpar/\kperp$ this is achieved if $m_3/m_1 \sim \epsilon$. 

 We let then $m_3 = \epsilon \tilde m_3$ and $\kperp =  \tilde\kperp/\epsilon$ so that $
\kperp m_3 \sim \kpar m_1.$  The asymptotic method that we use below  has been employed previously \cite{Depassier2015} for a thin film in the absence of current and DMI.

 Since the LLG equation implies that the  modulus of the magnetization is constant, the condition $|\vec m|=1$ together with the scaling implies
\be
m_1 ^2 + m_2^2  + \epsilon^2 \tilde m_3^2 = 1.
\ee

We search  for a solution of the LLG equation perturbatively. Letting 
\be 
\begin{split}
m_i &= m_i^0 + \epsilon m_i^1 +\ldots,  \text{ for } i=1,2,  \\
 \tilde m_3 &=   m_3^0 + \eps m_3^1 + \ldots, \\
 \htot &=  \vec h^0 + \eps \vec  h^1 +  \ldots,
 \end{split}
 \ee
we find that the leading order components satisfy
\begin{subequations}\label{cero}
\begin{align}
&(m_1^0)^2 + (m_2^0)^2 = 1, \label{inplane} \\
&\frac{\partial m_1^0}{\partial \tau} = - m_2^0 \,  h_3^0,  \label{uno}\\
&\frac{\partial m_2^0}{\partial \tau} = + m_1^0\,  h_3^0, \label{dos}\\
& 0 = m_2^0 h_1^0 - m_1^0 h_2^0 + \alpha \left( m_1^0 \frac{\partial m_2^0}{\partial \tau} - m_2^0 \frac{\partial m_1^0}{\partial \tau}  \right). \label{tres}
\end{align}
\end{subequations}
Using (\ref{inplane},\ref{uno},\ref{dos}) in (\ref{tres}) we obtain
\be \label{h3}
h_3^0 = \frac{1}{\alpha}  ( m_1^0 \, h_2^0(\vec m^0) - m_2^0\, h_1^0(\vec m^0) ).
\ee
On account of (\ref{inplane}) we express the in-plane  leading order magnetization as
$$
m_1^0 = \cos\varphi, \qquad m_2^0 = \sin\varphi,
$$
Notice then that equations (\ref{uno}) and (\ref{dos}) are equivalent, $\dot\varphi = h_3^0$. 
Replacing  (\ref{h3}) we obtain the time evolution equation
\be
\label{ecfin}
\alpha \frac{\partial \varphi}{\partial \tau}  =  h_2^0 \cos\varphi -   h_1^0 \sin\varphi \equiv h_{\varphi}^0.
\ee
The perpendicular total field satisfies 
\be\label{hefin}
h_3^0  =  \frac{\partial \varphi}{\partial \tau} .
\ee

Finally we need to calculate the leading order expansion for the total field. The scaling for the magnetization leads to the following expansion for $\htot$:
\begin{subequations}\label{fields}
\begin{align} 
h_1^0  &= h_a + \frac{\kpar}{2} {m_1^0}_{\xi\xi} + \kpar m_1^0 - \beta U {m_1^0}_{\xi},\\
h_2^0 &= \frac{\kpar}{2} {m_2^0}_{\xi\xi} - \beta U {m_2^0}_{\xi},\\
h_3^0  &= -  \tilde \kperp   m_3^0 - d {m_2^0}_{\xi} - U ( m_1^0  {m_2^0}_{\xi} -  m_2^0  {m_1^0}_{\xi}).
\end{align}
\end{subequations}

Replacing the leading order expansion for the field,  Eq. (\ref{fields}),  in (\ref{ecfin},\ref{hefin}), we obtain that the time evolution of  the domain wall is governed by
\be\label{rd}
\alpha \dot \varphi  =  - \beta U \varphi_{\xi} + \frac{1}{2}  \kpar \varphi_{\xi\xi}  -  h_a \sin\varphi -  \kpar \sin\varphi \cos\varphi,
\ee
and the perpendicular component of the magnetization is calculated from
\be\label{m3}
m_3^0 = - \frac{1}{\tilde \kperp} (  U \varphi_{\xi} + d  \varphi_{\xi}  \cos\varphi + \dot\varphi).
\ee

Equations ( \ref{rd})  and (\ref{m3}) constitute the central result of this work, the analysis of which we give below. 

\section{Front dynamics}

\begin{figure}
\centering
\includegraphics[width=0.5 \textwidth]{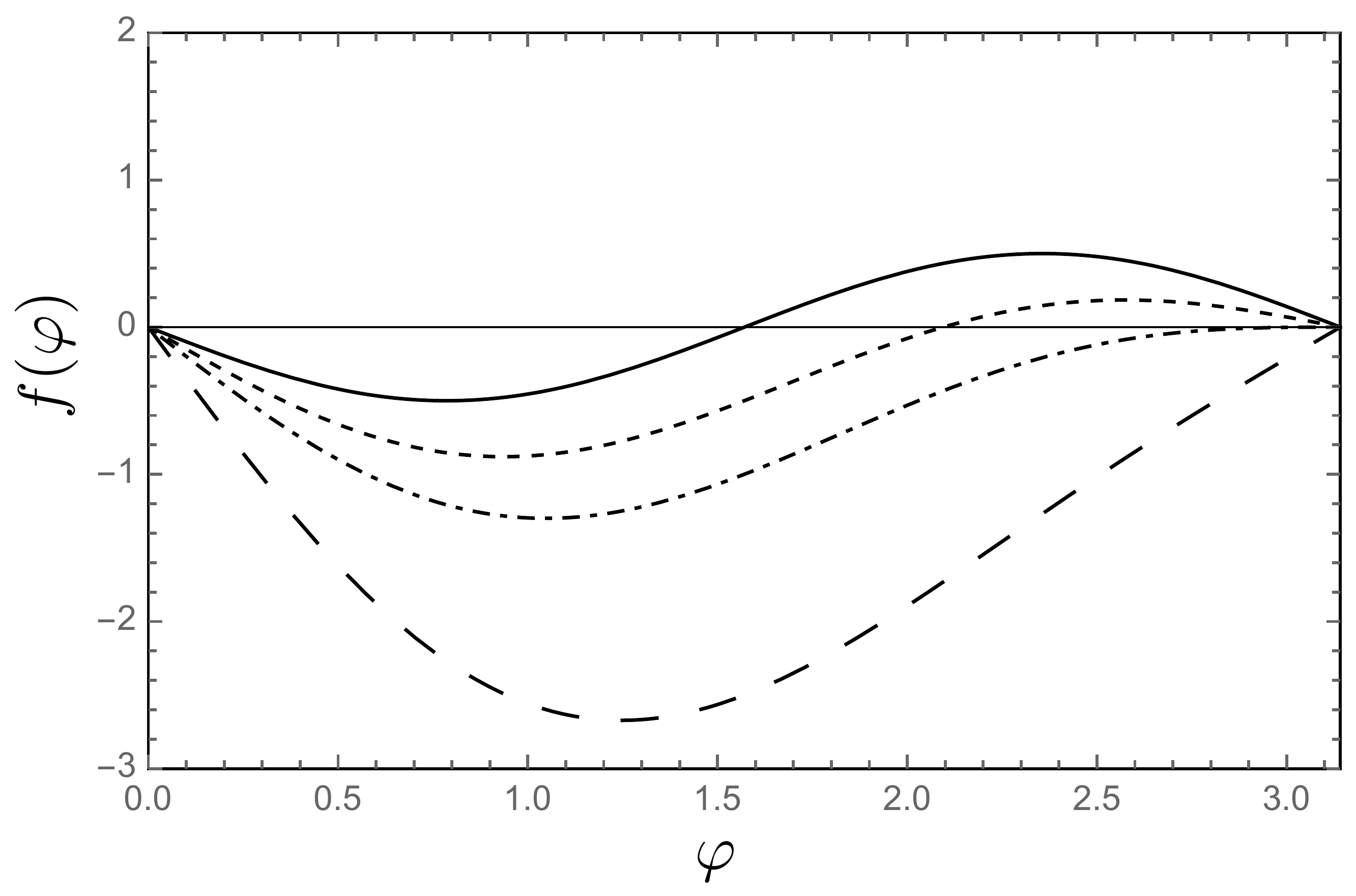}
\caption{Reaction term $f(\varphi)$ for $\kpar=1$ and different values of the applied field. The solid line shows the profile for the static solution $h_a=0$ for which $\varphi =0$ and $\varphi=\pi$ are equally stable. As the magnetic field increases $\varphi =0$ augments its stability while $\varphi=\pi$ gradually loses stability until it becomes  unstable at sufficiently large applied field. In the figure the short dashed line, the dot-dashed line and the long dashed line correspond to dimensionless fields $h_a= 0.5, 1 \text{ and } 2.5 $ respectively}
 \label{fun}
\end{figure}

Equation (\ref{rd}) is the well studied one dimensional reaction diffusion equation with   the reaction term 
\be\label{reaction}
f(\varphi)  \equiv  -\sin\varphi (h_a + \kpar \cos\varphi).
\ee
The dynamics of a domain wall in a nanotube including DMI has been recently studied \cite{Goussev2016} and the dynamics is also governed by a reaction diffusion equation with a reaction term dependent on the DMI.  For the nanotubes the speed  and profile can be solved analytically only in the case of vanishing applied field.  When an external field  is applied an exact solution cannot be constructed,   An expansion for  small  applied fields  shows the effect of DMI on the speed.  For nanotubes the  small radial component of the magnetization  has not been calculated. 
 In the present problem the speed is not affected by DMI, the speed, profile  and perpendicular magnetization can be calculated explicitly in the presence of current and magnetic field.  This allows to  show that there is a transition   to a  pulled or KPP \cite{KPP, VanSaarloos89}  regime at larger applied field as we show below.
 Notice that  the reaction term  Eq. (\ref{reaction})  is odd in $\varphi$ so that if $\varphi(\xi,\tau)$ is a solution to (\ref{rd}) then  $-\varphi(\xi,\tau)$ is also a solution, so we concentrate only on positive solutions in the interval $0\le \varphi \le \pi.$
More importantly the reaction term changes from bistable to monostable as the applied field increases which is important for the dynamics. In Fig. \ref{fun} we have plottted the reaction term for $\kpar=1$ and different values of the applied field. For  general positive values of $\kpar$  the transition from a bistable reaction function to a monostable one occurs at $h_a = \kpar$. 

The evolution of an initial condition $\varphi(x,0)$ under Eq. (\ref{rd}) has been fully studied for all types of reaction terms. It is known that in the bistable regime there is a unique traveling wave solution joining the two stable states, and in the monostable state, here $h_a>\kpar$, there is a continuum of traveling wave solutions 
\cite{Fife1977, Aronson1978}. 
In the bistable case  a suitable initial condition evolves into the unique traveling wave \cite{Fife1977, Aronson1978}  and in the monostable case it evolves into the traveling wave of minimal speed.  The problem in this regime is to determine  this   minimal speed.  The initial condition in the present case is  a static HH or TT domain wall which satisfies  the hypothesis of \cite{Fife1977}. For the sake of completeness we recall the main facts needed to determine the speed of the domain wall. 

It is convenient to introduce a change of variables in order to apply directly the standard mathematical results. 
Going to the moving frame $\eta = \xi - (\beta U/\alpha)  \tau$  and introducing the new independent variable $\phi = 1 - \varphi/\pi$,  Eq.(\ref{rd}) becomes
\be\label{rd2}
\alpha \frac{\partial \phi}{\partial \tau}  =  D \phi_{\eta\eta}   + F(\phi),  \text{   with   }  F(\phi)= \frac{  \sin \pi\phi}{\pi} ( h_a -  \kpar  \cos\pi \phi)
\ee
and $ D =   \kpar/2 $.
In these new variables, the reaction term $F(\phi)$ satisfies $F(0) =F(1) = 0$ and  for $h_a< \kpar$ is  bistable, that is,  $F<0$ in $(0,a)$, $F>0$ in $(a,1)$, $0<a<1$ . If $h_a \ge \kpar$ it is monostable, i.e.,  $F>0$ in $(0,1)$.  

This  equation has the exact solution 
\be\label{solex}
\phi_\sigma(q) = \frac{2}{\pi} \arctan\left(\text{exp}(\sigma \sqrt{2}\, q)\right), \text{ with  }  \qquad c_{\sigma} = -\sigma \frac{h_a}{ \sqrt{2}\, \alpha}.
\ee
where $q= \eta- c \tau,$ and $\sigma = \pm 1$. 
This exact solution is the unique solution in the bistable regime. In the monostable regime it is one of a continuum of solutions, it will be the solution to which a perturbation of the static state converges only if it is the front of minimal speed.  

The standard theory \cite{KPP, Aronson1978} guarantees that  in the monostable regime suitable initial conditions will evolve into a traveling wave  of minimal speed  $\phi(\eta-  c_{\text{min}} \tau)$ and the  minimal speed $c_{\text{min}}$  satisfies
 $$
 c_{\text{min}} \ge \frac{2 }{\alpha}  \sqrt{ D F'(0)}  \equiv c_{\text{\tiny{KPP}}}.
 $$

We will show that the exact solution Eq. (\ref{solex}) is the  solution selected by the dynamics  for $0 < h_a \le 2 \kpar.$ 
For $  h_a > 2 \kpar$ the minimal speed is the KPP value.  The minimal speed  for a monotonic front of (\ref{rd2}) satisfies the variational characterization \cite{BDPRL1996}
\be\label{var}
c^2 = \sup_g \frac{ 2 D}{\alpha^2} \frac{ \int_0^1 F(u) g(u) du}{ \int_0^1  g^2(u) h^{-1} (u)  du},
\ee
where $g(u)$ is an arbitrary positive function such that $h(u) = - g'(u) >0$ and such that the integrals  in  (\ref{var}) converge. In the case where an exact solution exists one can find the optimizing function $g(u)$, say $\hat g(u)$ for which the equal sign holds in  (\ref{var}).
In effect, choosing as a trial function
$$
g(u) = \left[ \tan\left( \frac{\pi u}{2} \right)\right] ^{- h_a/\kpar}
$$
we obtain
$$
c^2 \ge \frac{h_a^2}{2 \alpha^2} \quad\text{for}\quad -2 \le h_a/\kpar \le 2.
$$
This coincides with the exact solution so this is the optimizing trial function $\hat g$ and this is the minimal speed for $h_a \le 2 \kpar.$ The exact solution (\ref{solex}) is the profile in this regime. For larger applied fields the speed, in the moving coordinate reference frame is $c_{\text{\tiny{KPP}}}$. The magnetization profile for $h_a > 2 \kpar$ cannot be found analytically except at the transition point $h_a = 2\kpar$ but it shares the qualitative features of the exact solution.

\begin{figure}[H]
\centering
\includegraphics[width=0.5 \textwidth]{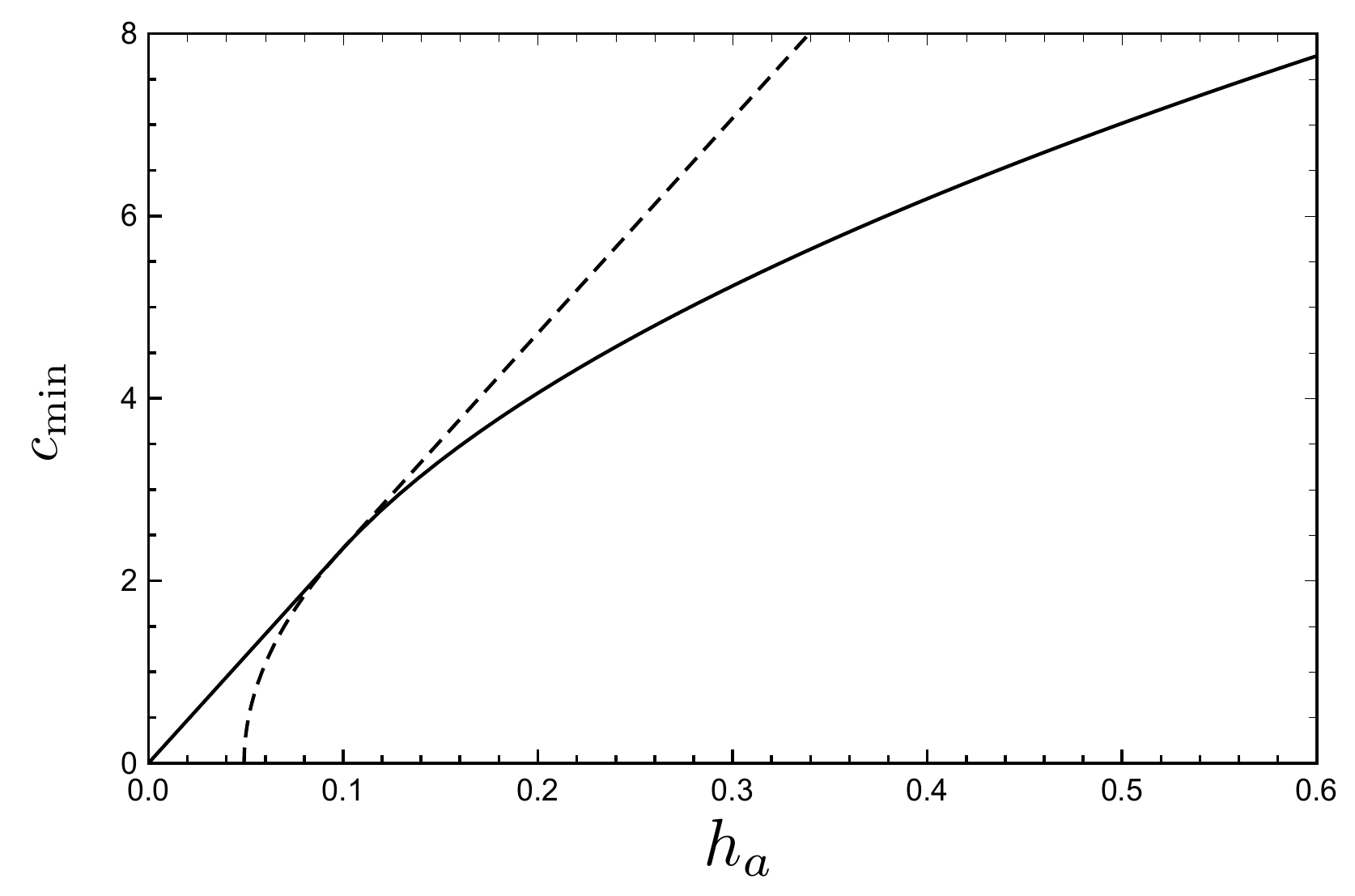}
\caption{
The solid line shows the domain wall speed as a function of the magnetic field for $\alpha=0.03, \kpar=0.049$. For $h_a > 2 \kpar$ the domain wall is of KPP type, for $h_a\le 2 \kpar$ it is a  pushed front.
}
 \label{Figspeed}
\end{figure}

In Fig. \ref{Figspeed}  we show the absolute value of the speed in the moving frame as a function of the applied field. 
We choose as parameters those of  sample A3 of \cite{Thevenard2012}, namely $M_s = 36   \text{ kA m}^{-1}, K_u = 40 \text{ J m}^{-3}, K_d  = 7415 \text{ J m}^{-3},  C_{ex} = 2 \times 10^{-13}   \text{ J m}^{-1}$ and  the Gilbert constant $\alpha= 0.03$.  The solid line shows the speed for all values of the applied field. The transition to the KPP regime occurs at   $h_a = 0.098$ in dimensionless units. 
For $0<h_a \le 0.098$ the speed is that given by the exact solution Eq. (\ref{solex}) whereas for $h_a> 0.098$ the speed is the $c_{\text{\tiny{KPP}}}$ value. The dot-dashed line shows the exact speed  Eq. (\ref{solex}) which is not the speed of the domain wall in that parameter regime. Likewise the dashed line is the $c_{\text{\tiny{KPP}}}$ value, in the region where it  is not the selected speed.

In summary, going back to the original independent variable $\varphi$ and the laboratory frame, we find that the speed is given by
$$
v_{\sigma} =  \left \{\begin{array}{ll} 
 \frac{\beta U}{\alpha} -   \frac{\sigma h_a}{\sqrt{2}\, \alpha} \qquad \qquad \qquad \qquad  & {\rm if} \,  h_a \le 2 \kpar, \\
 \frac{\beta U}{\alpha} -  \frac{\sigma \sqrt{2 \kpar}}{\alpha}  \sqrt{h_a-\kpar}  & {\rm if} \,  h_a > 2 \kpar. \end{array} \right.
$$

In the first case the domain wall profile is determined by the analytic solution (\ref{solex}). In the original laboratory coordinates 
$$
\varphi_{\sigma} = \pi - \pi \phi_{\sigma} = \pi \phi_{-\sigma} = 2 \arctan\left(\text{exp}(- \sigma \sqrt{2} ( \xi-v_{\sigma}\tau )\right),
$$
so that 
$$
m_1 = \sigma \tanh(\sqrt{2}(\xi - v_{\sigma} \tau) ),\qquad \qquad m_2 =  \sech(\sqrt{2}(\xi - v_{\sigma} \tau) ),
$$
and 
$$
m_3 = \sqrt{2}\sigma  \sech(\sqrt{2}(\xi - v_{\sigma} \tau) ) \left(  \frac{\sigma h_a}{\sqrt{2}\, \alpha} + U( 1 - \frac{\beta}{\alpha}) + \sigma d  \tanh(\sqrt{2}(\xi - v_{\sigma} \tau) \right).
$$
The TT (HH)  solution corresponds to  $\sigma=1 (-1)$ respectively.

The solution joining $-\pi$ to $0$ differs in chirality, the solution is analogous. 

\begin{figure}[h!]
\centering
\includegraphics[width=0.5 \textwidth]{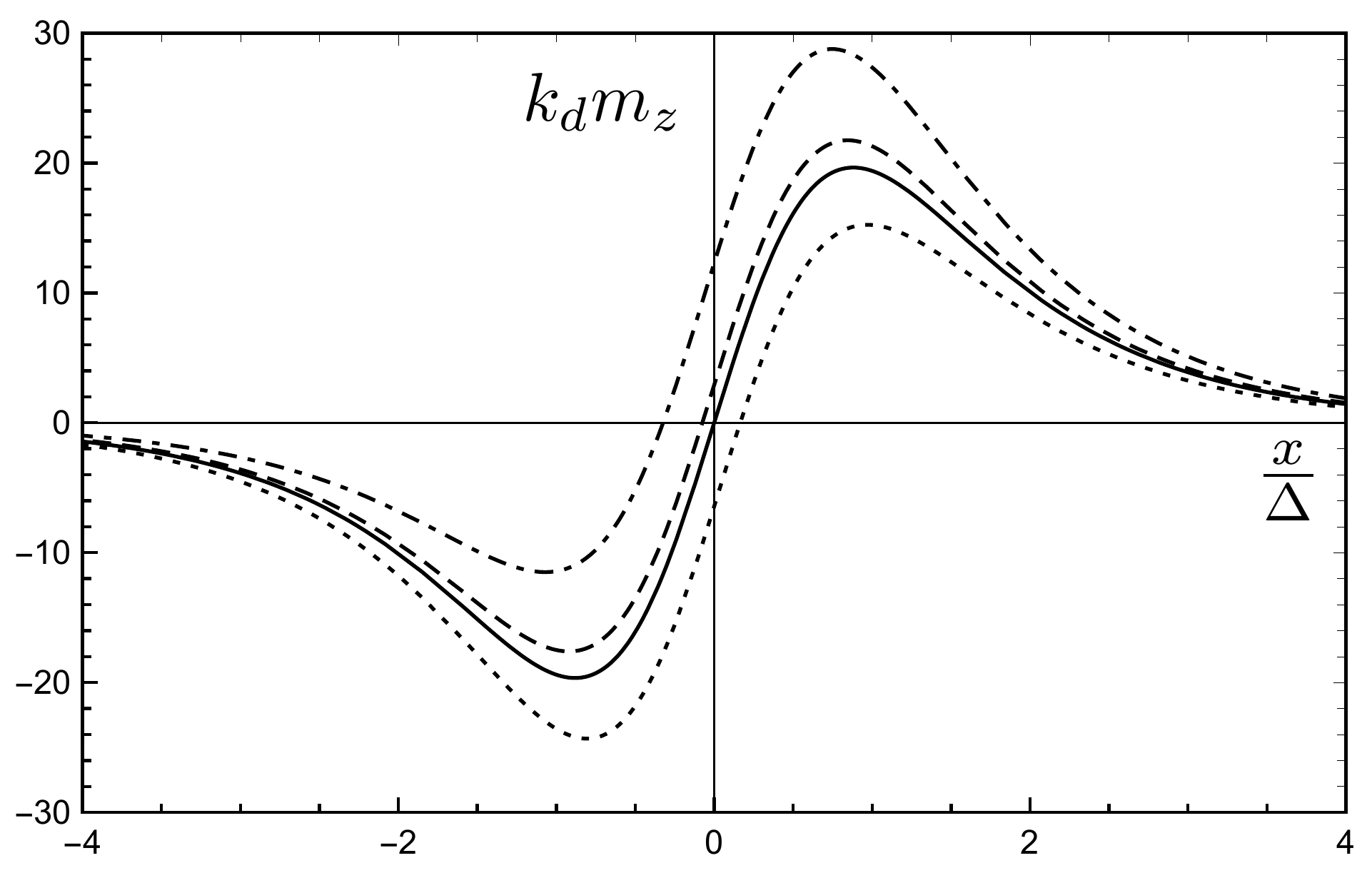}
\caption{The solid line shows the static out of plane magnetization. The  dashed line shows the profile for vanishing current and applied field $H_a = 0.088 M_s.$ The dotted and dot-dashed lines show the profile for the same value of the magnetic field and for $u= \pm 400$ ms$^{-1}$  respectively.}
 \label{fig1}
\end{figure}

The speed and in plane components of the magnetization  are unaffected by the DMI in this approximation, the out of plane component is distorted both by the applied field an current as well as the DMI. In Fig. \ref{fig1} we have plotted the out of plane magnetization component  for  different values of the current and magnetic field.  
We use  the same material   parameters as in Fig. 2 and  the values  $D = 1.6 \times 10^{-3}$J m$^{-2},  \beta=10 \alpha$.

\section{Summary}

We studied the dynamics of an in-plane magnetized thin film including DMI  under applied current and field acting along the easy axis when the perpendicular anisotropy is large.  In this limit the 
out of plane magnetization is slaved to the in-plane components,  the dynamics of which is governed by a reaction diffusion equation. 
Reaction diffusion dynamics is also encountered in the study of thin nanotubes however the role played by DMI in the dynamics is different for thin films. 
As mentioned in the introduction previous studies addressed the case of negligible $\kperp$  and the limit of small DMI. Here we have considered  the limit of large $\kperp$ and  find significant differences  in the  dynamics. For negligible $\kperp$ \cite{Tretiakov2010}  the domain wall width and speed   depend on the DMI parameter and  the magnetization spins. For small DMI \cite{Kravchuk2014} and perpendicular anisotropy $\kperp$  of order one the DMI does not affect the speed nor the  in-plane magnetization profile in agreement with the present results.  The effect of small DMI is the introduction of an asymmetric distortion in the out of plane component and a shift in the Walker field. In the case studied in this work the Walker field is not observed;  instead,  for sufficiently  large applied field the domain wall slows down due to a change in the nature of the domain wall which goes from bistability to monostability and enters the so called KPP or pulled regime.  Increasing the current does not lead to this change of behavior.  Previous numerical work has reported the slowdown of the domain wall in thin films with very large axis anisotropy beyond a critical applied field due to spin wave emission \cite{Wieser2010,Wang2012,Wang2014}. Here we find the same qualitative feature, further study is required to obtain quantitative agreement with the numerical results and understand if the transition form pushed to pulled fronts is due to the emission of spin waves. The reaction diffusion equation filters out and gives no information on the emission of spin waves. In addition we were able to obtain a simple explicit expression for the out of plane magnetization showing the effect of the DMI and the motion on the profile showing an asymmetric distortion of the out of plane component when the domain wall moves. The perturbation method that we have used may prove useful to study other configurations where a N\'eel wall is preferred. 

\section{Acknowledgments}

We acknowledge helpful discussions with E. Stockmeyer. This work was partially supported by Fondecyt (Chile) projects 114--1155
 and 116--0856.


%

 \end{document}